\documentclass[aps,prl,twocolumn,nobibnotes,superscriptaddress,citeautoscript]{revtex4-1}
\usepackage{amsmath}
\usepackage{amssymb}
\usepackage[pdftex]{graphicx}
\usepackage{wrapfig}
\usepackage{subfigure} 
\usepackage{color}

\newcommand{\supplementary}{supplemental materials(SM)}
\newcommand{\supacronym}{SM}

\begin{document}
\DeclareGraphicsExtensions{.pdf}
\title{Striped spin liquid crystal ground state instability of kagome antiferromagnets}
\author{Bryan K. Clark}
\thanks{These authors contributed equally.}
\affiliation{Station Q, Microsoft Research, Santa Barbara, CA 93106, USA}
\affiliation{Princeton University and Princeton Center for Theoretical Science, Department of Physics, Princeton, NJ 08544}
\author{Jesse M. Kinder}
\thanks{These authors contributed equally.}
\affiliation{Case Western Reserve University, Department of Physics, Cleveland, OH 44106}
\affiliation{Cornell University, Department of Chemistry, Ithaca, NY 14853}
\author{Eric Neuscamman}
\affiliation{University of California Berkeley, Department of Chemistry, Berkeley, CA 94720}
\affiliation{Cornell University, Department of Chemistry, Ithaca, NY 14853}
\author{Garnet Kin-Lic Chan}
\affiliation{Princeton University, Department of Chemistry, Princeton, NJ 08544}
\affiliation{Cornell University, Department of Chemistry, Ithaca, NY 14853}
\author{Michael J. Lawler}
\email{mlawler@binghamton.edu}
\affiliation{Department of Physics, Binghamton University, Binghamton, NY 13902-6000} 
\affiliation{Department of Physics, Cornell University, Ithaca, NY 14853} 
\date{\today}

\begin{abstract}
The Dirac spin liquid ground state of the spin 1/2 Heisenberg kagome antiferromagnet has potential instabilities \cite{Hastings2000,Hermele2005,Ran2007,Hermele2008}. This has been suggested as the reason why it is not strongly supported in large-scale numerical calculations\cite{Yan2011}. However, previous attempts to observe these instabilities have failed. We report on the discovery of a projected BCS state with lower energy than the projected Dirac spin liquid state which provides new insight into the stability of the ground state of the kagome antiferromagnet.  The new state has three remarkable features. First, it breaks both spatial symmetry in an unusual way that may leave spinons deconfined along one direction. Second, it breaks the U(1) gauge symmetry down to $Z_2$.   Third, it has the spatial symmetry of a previously proposed ``monopole'' suggesting that it is an instability of the Dirac spin liquid. The state described herein also shares a remarkable similarity to the distortion of the kagome lattice observed at low Zn concentrations in Zn-Paratacamite suggesting it may already be realized in these materials.
\end{abstract}

\maketitle

Many potential ground states have been suggested for the spin 1/2 kagome Heisenberg antiferromagnet;  these include magnetic ordering in one of several spin patterns\cite{Chubukov1992}, a valence bond crystal with a 36 site unit cell\cite{Marston1991} or a 12 site unit cell\cite{Yan2011}, a chiral spin liquid\cite{Marston1991,Messio2011}, several kinds of $Z_2$ spin liquids\cite{sachdev1992,Lu2011} and an algebraic spin liquid\cite{Hastings2000,Ran2007}.  Amongst algebraic spin liquids, it was shown \cite{Ran2007} that the $U(1)$-Dirac state had lowest variational energy.  This latter state is characterized by its graphene-like Dirac fermionic spinons interacting with a Maxwell-like photon that characterizes singlet excitations.  Many of the other suggestions for the kagome antiferromagnet ground state can be characterized as instabilities of this Dirac spin liquid; these instabilities have been catalogued in ref. \onlinecite{Hermele2008}.  However, despite the many instabilities, all variational studies we are aware of that  include the $U(1)$-Dirac state for some choice of variational parameters have found it remarkably stable\cite{Hermele2005,Nogueira2005,Ran2007,Hermele2008,Iqbal2011b}. 

Recently, large scale DMRG calculations\cite{Yan2011,Depenbrock2012,Jiang2012} have produced strong numerical evidence that the true ground state of the kagome antiferromagnet is a $Z_2$ spin liquid.  In particular, they have found the lowest known variational state with energies comparable to exact diagonalization on small clusters, a gap to all excitations and topological entanglement entropy expected of a $Z_2$ spin liquid. These results are quite surprising in light of prior studies of the U(1)-Dirac state and suggest that it is unstable.

Perhaps the simplest explanation of the DMRG results, as proposed in Ref. \onlinecite{Lu2011}, is that the Dirac-like spinons pair up and form a superfluid-like $Z_2$ spin liquid. With this goal in mind, these authors catalog 20 such possible states and give explicit instructions for constructing 14 of them.  Unfortunately, despite searches for such a state\cite{Iqbal2011b}, no $Z_2$ state has been shown to be energetically favorable to the  $U(1)$ Dirac state at the level of variational projected BCS wave functions. 

\begin{figure}[t]
\includegraphics[width=0.4\textwidth]{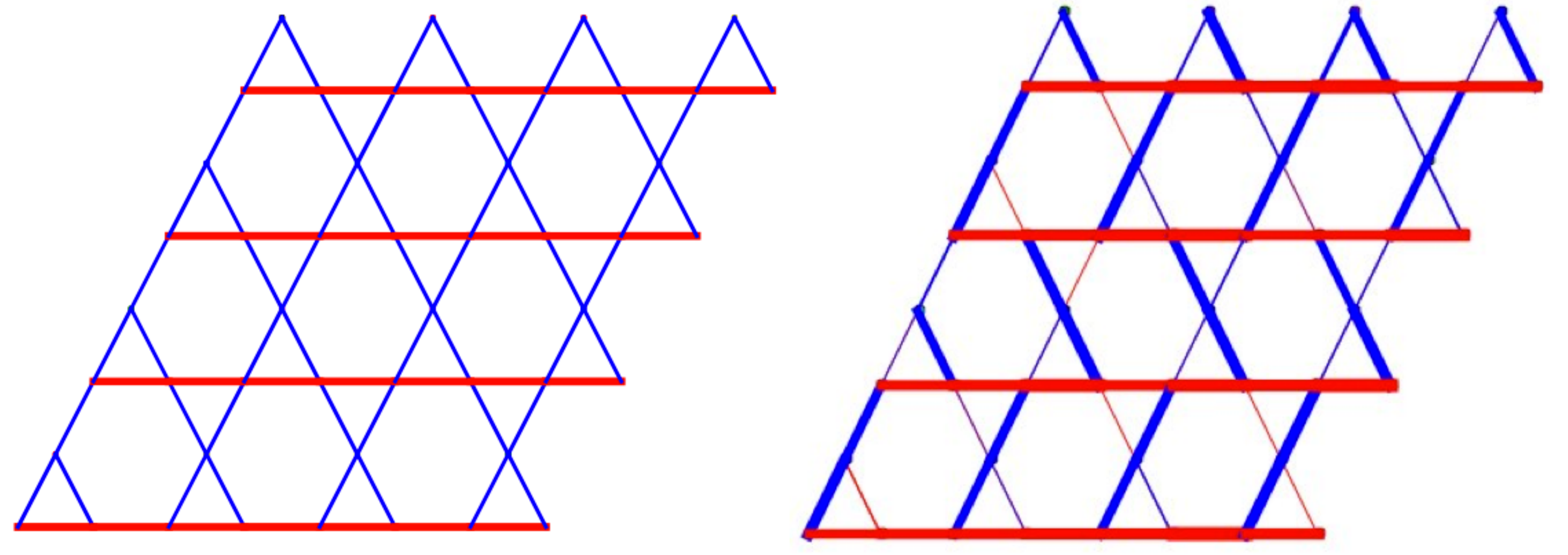}
\caption{The n.n. $\vec s \cdot \vec s $ correlations for the projected Dirac Spin liquid state with twisted boundary conditions (left) and
the lowest energy projected mean field state (right).  The color
indicates sign and the linewidth measures the magnitude. }
\label{fig:S.S}
\end{figure}

In this letter, we revisit the projected BCS (PBCS) variational wave function problem on the kagome lattice and have discovered an instability of the Dirac state.  Our approach, building on Refs. \onlinecite{Neuscamman2012,Clark2011}, is to optimize over the entire set of time-reversal invariant projected BCS wave functions which are parameterized by a symmetric pairing matrix $\phi_{ij}$.  These states
encompass the U(1) and Z2 spin liquids, general valence bond solids, and a number of other instabilities. We find states with lower energy than the Dirac state on clusters of up to 192 sites.  On the 48 site cluster in particular, a carefully optimized wave function, as shown in Fig. \ref{fig:S.S}, demonstrates that the new state breaks lattice symmetries, doubling the unit cell with a wave vector that connects the two Dirac nodes. However it preserves  a $C_2$ rotational symmetry about several lattice points giving it a distinct one dimensional character. It also lies very close to the Dirac state with similar fluxes through the hexagons and bow ties. For these reasons, we believe it should belong to one of the leading instabilities proposed in Ref. \onlinecite{Hermele2008}, and have discovered that the most likely candidate is the time-reversal symmetric ``$w$-monopole''. We also find that this state cannot be obtained by optimizing within the class of wave functions that do not admit pairing and therefore likely has $Z_2$ global gauge symmetry. For this reason and because of its $C_2$ symmetry and one dimensional character, we have dubbed this state a $Z_2$ striped (or smectic) spin liquid crystal.  Finally, we discuss a connection between our state and Zn-Paratacamite with Zn concentrations $x$ below $x=1/3$ whose observed distorted kagome layers\cite{SHLee2007} has the same symmetry as Fig. \ref{fig:S.S}(b).

\begin{figure}[t]
\subfigure[]{\includegraphics[width=0.23\textwidth]{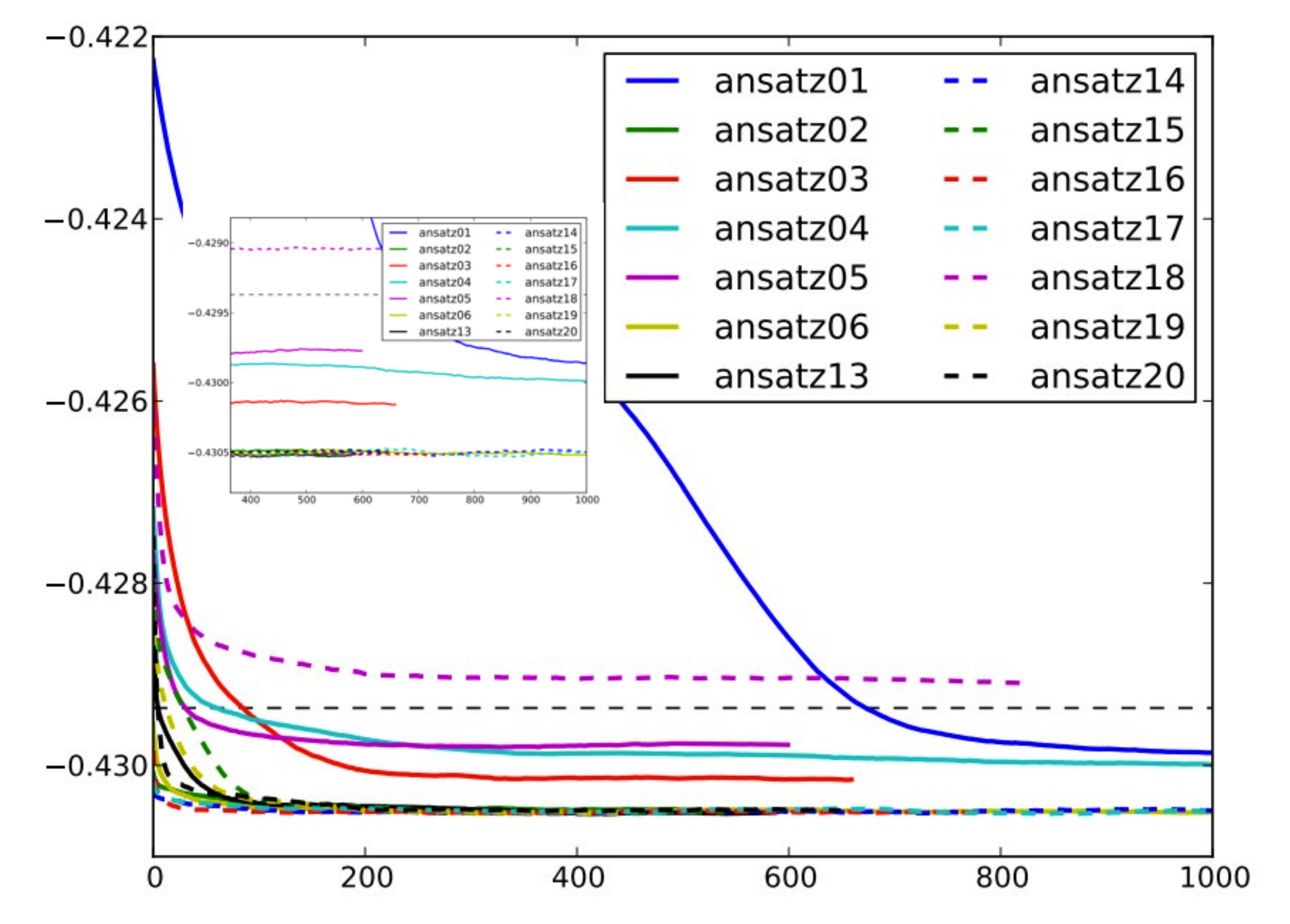}}{}
\subfigure[]{\includegraphics[width=0.2\textwidth]{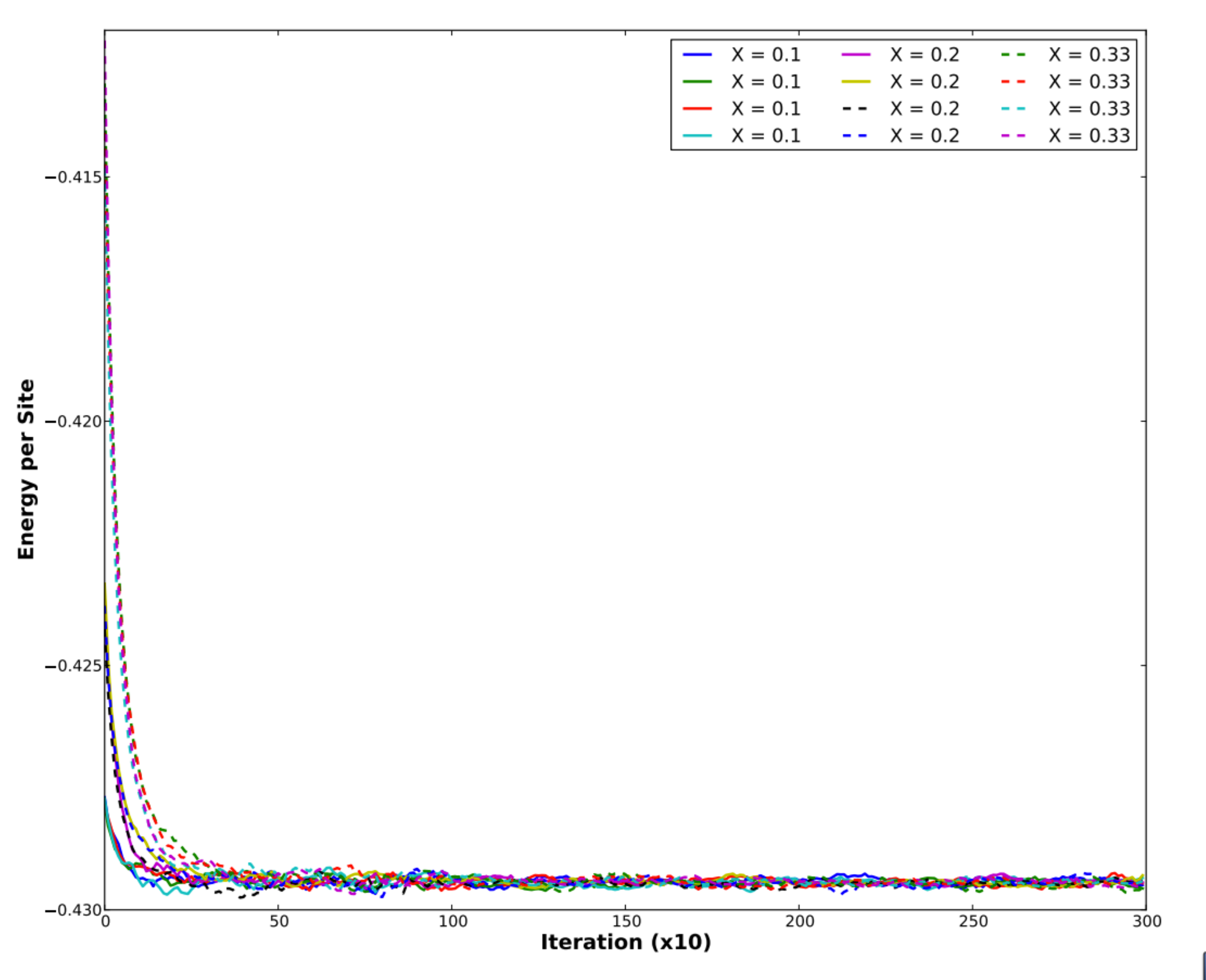}}{}
%\includegraphics[width=0.5\textwidth]{assym.png}
%\includegraphics[width=0.4\textwidth]{goodAssym.pdf}
%BRYANREMOVE \subfigure[]{\includegraphics[width=0.22\textwidth]{PF_best.pdf}}{}
%BRYANREMOVE \subfigure[]{\includegraphics[width=0.22\textwidth]{PF_absDeviation.pdf}}{}
\caption{(a): Prototypical stochastic optimization traces starting from the 14 distinct PSG ansatz defined in Table II of Ref. \onlinecite{Lu2011} and optimizing the pairing function $\phi$.  
Inset shows magnified version. (b): Optimization of 
only single particle orbitals starting from perturbed Dirac states where $X$ labels the strength of the perturbation. 
%BRYANREMOVE(c):  Values of the pairing matrix for our optimal state $\phi(\vec r_{i\uparrow},\vec r_{j\downarrow})$. (d): Values of $|\phi| - \overline{|\phi|}$.  The color in (c) and (d) indicates the sign and the linewidth measures the magnitude.  
}
\label{fig:Opt}
\end{figure}

%\section{A Broken Symmetry State}
Our objective is to find the ground state of the nearest neighbor spin $1/2$ kagome antiferromagnet constrained to the set of Gutzwiller projected variational BCS wave functions 
\begin{equation}\label{eq:Gutzwiller}
  |\Psi\rangle = \hat{\mathcal P}_{S=1/2}|\Psi_0\rangle
\end{equation}
where $\Psi_0(R)=\langle R|\Psi_0\rangle=\det M(R)$, 
$M_{ij}(R) = \phi(\vec r_{i\uparrow},\vec r_{j\downarrow})$
is the BCS pairing amplitude  and $\hat P_{S=1/2}$ projects these states onto the physical singly occupied Hilbert space of spin wave functions.  This set of variational ansatz includes projected Slater determinants such as the $U(1)$ Dirac spin liquid \cite{Ran2007}. The variational search is performed by minimizing the ground state energy 
\begin{equation}
  E = \langle \Psi|\hat H|\Psi\rangle,\quad \hat H = \sum_{\langle ij\rangle}\hat S^\alpha_i\hat S^\alpha_j.
\end{equation}
of the Heisenberg spin Hamiltonian on the kagome lattice using as parameters all $n(n-1)/2$ real values of the pairing function
$\phi(\vec r_i, \vec r_j)$ (building on Refs. \onlinecite{Neuscamman2012,Clark2011}). To avoid only finding local minima in this energy landscape, we choose to start the optimization
in many qualitatively different starting points by
making use of the well developed projective symmetry group classification developed in the literature\cite{Wen2002, Ran2007, Hermele2008, Lu2011}. In particular, we will choose mean field parameters following Ref. \onlinecite{Lu2011} and derive a pairing matrix from these following \supplementary{} S-I and S-II. This allows us to start from 14 different spin liquid wave functions.  Fig. \ref{fig:Opt}(a) shows prototypical optimization traces of the energies for each spin liquid state on  the $4 \times 4$ lattice, labeled following Table II of Ref. \onlinecite{Lu2011}. Many states lie below the energy of the Dirac spin liquid and the minimal energy state we find is at $-0.430520 \pm 5\times 10^{-6}$ per site (see \supacronym{} S-III for its pairing matrix), significantly below the Dirac spin liquid energy of $−0.42938 \pm 4 \times 10^{-5}$ per site showing that the Dirac spin liquid is not the most stable projected mean field state. 

This is particularly surprising in light of previous studies that find the Dirac spin liquid to be stable against many instabilities\cite{Hastings2000,Ran2007,Hermele2008,Ma2008,Iqbal2011b}. Such calculations failed to find this state because they only considered short-range mean field Hamiltonians, a limitation we avoid by optimizing a general pairing function that allows both long-range hopping terms and spatial symmetry breaking in the fermion Hamiltonian. As shown in Fig. \ref{fig:Opt}(b) and discussed in more detail below, our approach in conjunction with enforcing $U(1)$ symmetry does not lower the energy. But, additionally allowing the $U(1)$ gauge symmetry to break down to $Z_2$ does lower the energy (left side of Fig. \ref{fig:Opt}). Consequently, the gauge symmetry breaking down to $Z_2$ is intricately linked to spatial symmetry breaking and also possibly long-range hopping/pairing terms in the Hamiltonian.

%\begin{figure}[h]
%\includegraphics[width=0.22\textwidth]{PF_best.pdf}
%\includegraphics[width=0.22\textwidth]{PF_absDeviation.pdf}
%\caption{Pairing functions for our optimal state for the n.n. bonds. The color
%indicates sign and the linewidth measures the magnitude.  Left:  Values of $\phi$.
%Right: Values of $|\phi| - \overline{|\phi|}$ }
%\label{fig:PairingFunction}
%\label{fig:PairingFunction}
%\end{figure}

%As in any projection scheme, the operator $\hat{\mathcal P}_{S=1/2}$ in Eq. \eqref{eq:Gutzwiller} performs a many-to-one mapping. This redundancy turns out to include an $SU(2)$ unitary rotation at each site\cite{Affleck1988,Dagotto1988}. Because a minimization algorithm can always get stuck in a local minimum of $E$, it will be important to start minimizing from many qualitatively different locations in the space of wave functions. To this end, we will make use of the well developed projective symmetry group classification developed in the literature\cite{Wen2002, Ran2007, Hermele2008, Lu2011}. In particular, we will follow Ref. \onlinecite{Lu2011} allowing us to start from 12 different spin liquid wave functions.
Let us then focus on the nature of our lowest energy state.  Most strikingly we find that the state breaks translational symmetry doubling the unit cell. Fig.~\ref{fig:S.S} (right) shows the deviation of the spin-spin correlation function on nearest neighbors from its average value, in contrast with the U(1) Dirac spin liquid shown in Fig.~\ref{fig:S.S} (left). The symmetry breaking seen in the Dirac spin liquid is a finite size artifact coming from twisted boundary conditions in one direction. In our optimized state we find breaking of the translational symmetry and a doubling of the unit cell leading us to conclude that the minimal projected mean field state is \emph{not} an isotropic spin liquid.  

%For a spin liquid, the pairing function $\phi$ should be symmetric under the symmetries of the lattice.  We see in Fig.~\ref{fig:Opt}(c) and (d)  that the pairing function for our optimized state is not symmetric showing a clear pattern in this representation. 

To further verify that there is not a nearby isotropic spin liquid state to our optimal state, 
we stochastically modify the pairing function, starting from a low energy state, in small increments searching for a true symmetric spin liquid.
In particular, we make a random stochastic change to the pairing function accepting it 
only if we have moved closer to a spin liquid state (defined by the deviations of the unprojected $\langle \vec S_i\cdot \vec S_j\rangle$).  In this way, we search for the ``closest'' spin liquid. When we run this procedure, the spin liquid 
state it approaches has the energy of the Dirac spin liquid 
(and hence is presumably the Dirac spin liquid). 
This leads us to believe that
their is no additional nearby isotropic spin liquid.

%\section{Closeness to the Dirac State}
%As numerical DMRG results strongly suggest that the ground state is a gapped spin liquid, 
%there is clear interest in understanding whether our lower energy state also suggests a spin liquid PMF of lower energy then the gapless Dirac state.  To examine this question, 
%we have searched for a nearby spin liquid to our symmetry-broken state. We do this by starting at a state close in energy to the symmetry-broken state and stochastically changing the pairing function to move it toward a more symmetric state.  We use the squared deviation from average of the n.n. unprojected $\vec s \cdot \vec s$ as a heurestic for assymetry and only walk uphill in this metric.  The `nearest' symmetric state we find is (at the energy of) the Dirac state. 
%Another way of understanding whether there are other nearby spin Later in this letter, 
%we will also quantify which spin liquid our SB state appears to best resemble.
%To further understand the energy of the symmetry breaking, we 
%\begin{figure}[h]
 %\includegraphics[width=0.5\textwidth]{fluxplot.pdf} 
%\caption{Flux results from the anomalous density matrix. These results show that our optimized state is indeed quite close to the U(1) Dirac state.}
%\label{fig:Flux}
%\end{figure}
Having found a broken symmetry state, we would like to now understand whether it retains characteristics reminiscent of any spin liquid phase. To make such a connection, it is useful to compute the anomalous density matrix from the unprojected wave function defined by the pairing matrix that corresponds to our lowest energy projected state. It is given by
\begin{equation}
  \rho_{ij} = \begin{pmatrix} -A_{ij}^* & B_{ij} \\ B_{ij}^* & A_{ij} \end{pmatrix}
\end{equation}
where 
\begin{equation}
A_{ij} = \frac{\langle\Psi_0|f_{i\downarrow}^\dagger f_{j\downarrow} |\Psi_0\rangle}{\langle\Psi_0|\Psi_0\rangle},\quad
B_{ij} =  \frac{\langle\Psi_0|f_{i\uparrow} f_{j\downarrow} |\Psi_0\rangle}{\langle\Psi_0|\Psi_0\rangle}
\end{equation} 
and transforms under an SU(2) gauge transformation, $\Psi\to{\bf G}\cdot\Psi$, like $\rho_{ij}\to {\bf G}_i\cdot\rho_{ij}\cdot{\bf G}_j^\dagger$ exactly like the mean fields in the corresponding slave particle theory (see Ref. \cite{WenText2004} or the \supacronym{} S-I). We can therefore use this matrix to study the projective symmetry properties of the obtained unprojected state $|\Psi_0\rangle$.

To study fluxes through the lattice, we can use the SU(2) matrix version of a ``phase'' variable
\begin{equation}
  {\bf W}_{ij} = -i \rho_{ij}/|\rho_{ij}|
\end{equation}
%BRYANREMOVE It has the property ${\bf W}_{ij}\cdot{\bf W}_{ij}^\dagger = {\bf W}_{ij}^\dagger \cdot{\bf W}_{ij} = {\bf I}$ and $\text{det}{\bf W}_{ij} = 1$ making it an $SU(2)$ matrix (the factor of $i$ makes the determinant positive). Together with the magnitude, the density matrix is then expressed in the form $\rho_{ij} = i|\rho_{ij}|{\bf W}_{ij}$.
that is an analog of the $U(1)$ phase $e^{ia_{ij}}$ we associate with ordinary electricity and magnetism on a lattice. Following Ref. \onlinecite{WenText2004}, the analog of flux through any loop on the lattice is therefore the product of this matrix around the loop
\begin{equation}
\Phi_{ijk\ldots l} = i^{N_{loop}}{\bf W}_{ij}\cdot{\bf W}_{jk}\cdot\ldots\cdot{\bf W}_{li}
\end{equation}
where $N_{loop}$ is the number of bonds $ij$ that form the loop. Unfortunately, this product is not gauge invariant,
%BRYANREMOVE .
%It transforms according to 
%\begin{equation}
%   \Phi_{ijk\ldots l}\to {\bf G}_i\cdot\Phi_{ijk\ldots l}\cdot {\bf G}_i^\dagger
%\end{equation}
%For this reason, the site $i$ is called the ``base point'' of the flux matrix. For an even number of sites, the flux matrix $\Phi_{ijk\ldots l}$ is also an element of $SU(2)$ and therefore has the form
%\begin{equation}
%  \Phi_{ijk\ldots l} = \cos(\theta/2){\bf I} + i\hat n\cdot\vec \tau \sin(\theta/2)
%\end{equation}
%where $\theta$ ranges from $0$ to $2\pi$. A gauge transformation therefore rotates the unit vector $\hat n$ but leaves $\theta$ unchanged.
%Hence
but for every loop of the lattice, we can define an angle $\theta$ associated with its flux through the trace of this matrix. Unlike a $U(1)$ flux, here $\theta=2\pi$ introduces a phase change of $-1$. The natural first loops to consider when characterizing the state $|\Omega\rangle$ are the nearest neighbor bow ties (product of two neighboring triangles) and hexagons (the trace of $\Phi_{ijk\ldots l}$ for an odd site loop vanishes by time reversal symmetry). These loops allow us to determine which of the four U(1) spin liquid states is closest to our best optimized state. The results are 
\begin{equation}
   \langle\theta_\text{hex}\rangle = (1.994 \pm 0.003)\pi, \langle\theta_{bow}\rangle = (0.010\pm .007)\pi
\end{equation}
where the error estimate is the standard deviation and $\langle\ldots\rangle$ denotes the average value of the flux over different hexagons or bow ties respectively.  These results are equivalent to nearly $\pi$ flux through the hexagon and $0$ flux through the bow tie or triangle in the traditional $U(1)$ description of flux. They show distinctly that the optimized state is very close to the $U(1)$ Dirac state as expected from energetic considerations but that because of the symmetry breaking shown in Fig.~\ref{fig:S.S} it is an instability. 
%\subsection{Relationship between energy and symmetry breaking}

\begin{figure}[t]
\subfigure[]{ \includegraphics[width=0.2\textwidth]{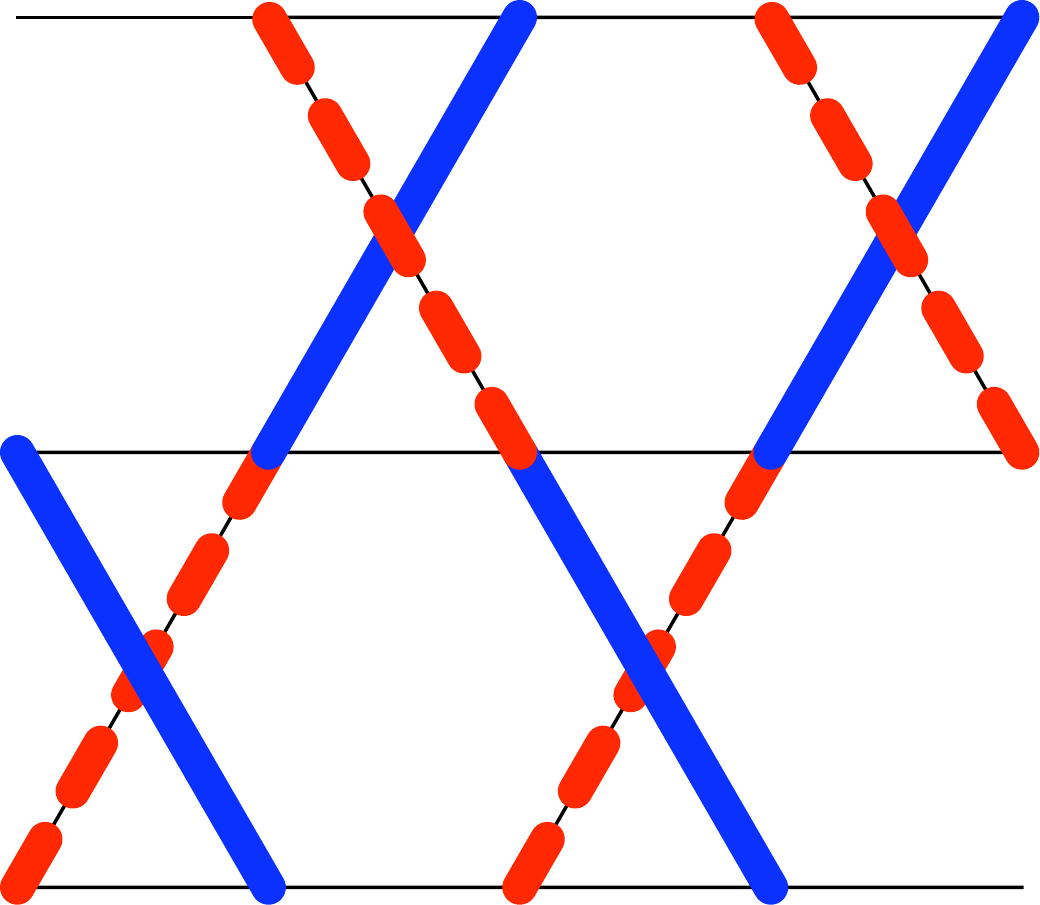}}
\subfigure[]{ \includegraphics[width=0.2\textwidth]{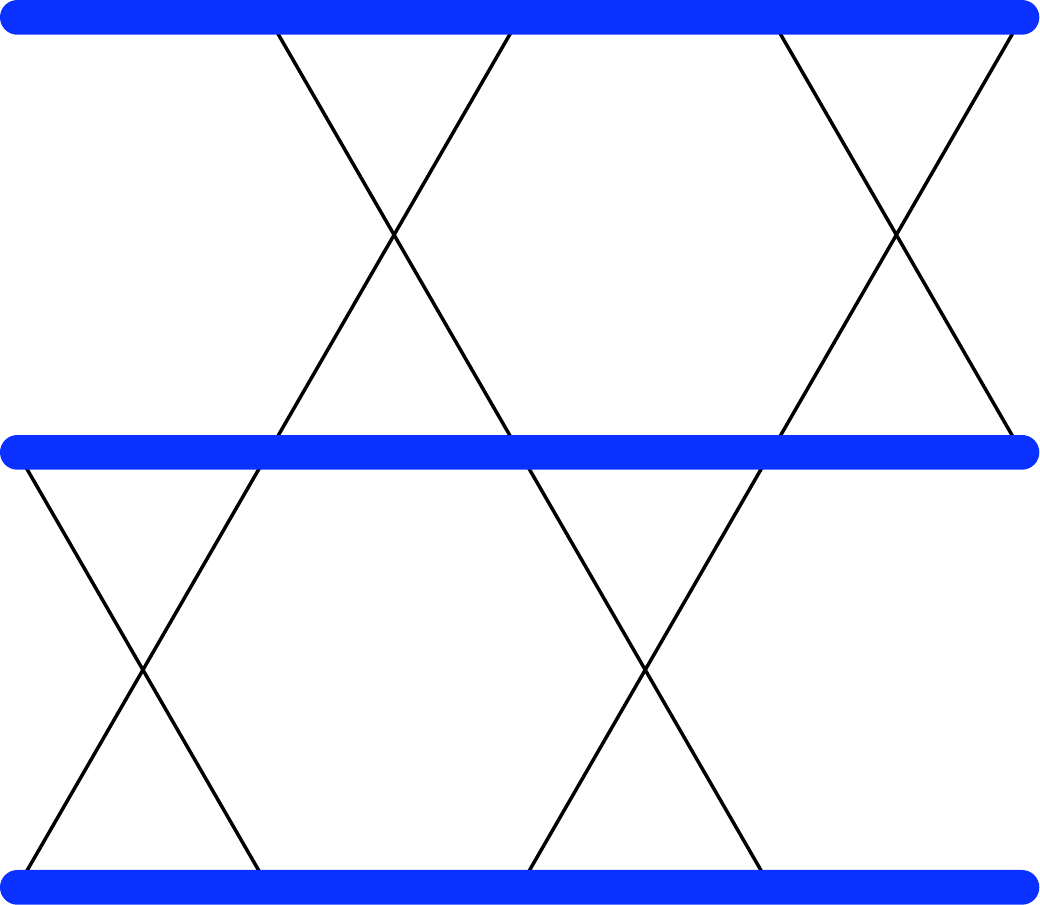}}
\subfigure[]{ \includegraphics[width=0.23\textwidth]{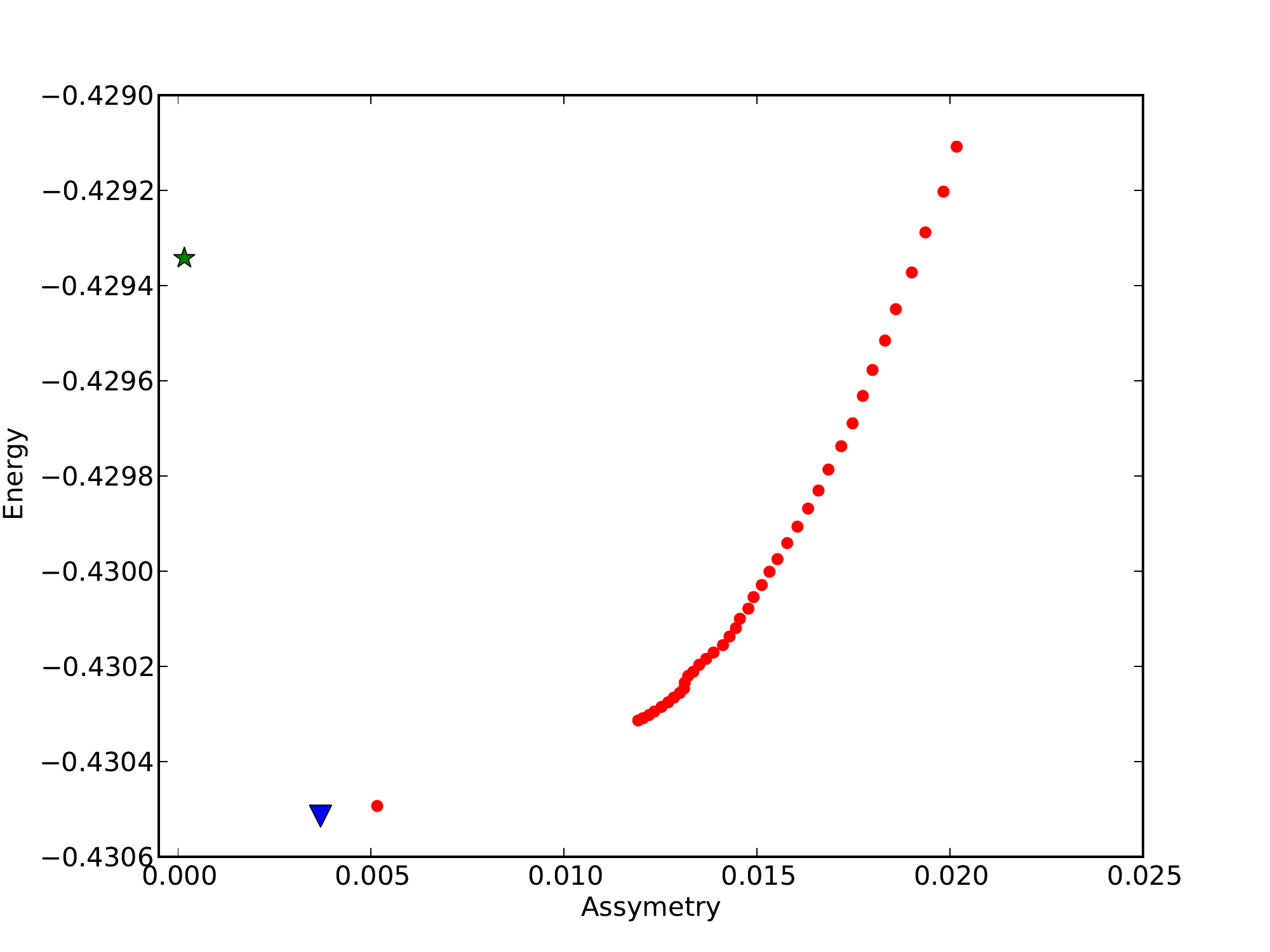}}
\subfigure[]{ \includegraphics[width=0.23\textwidth]{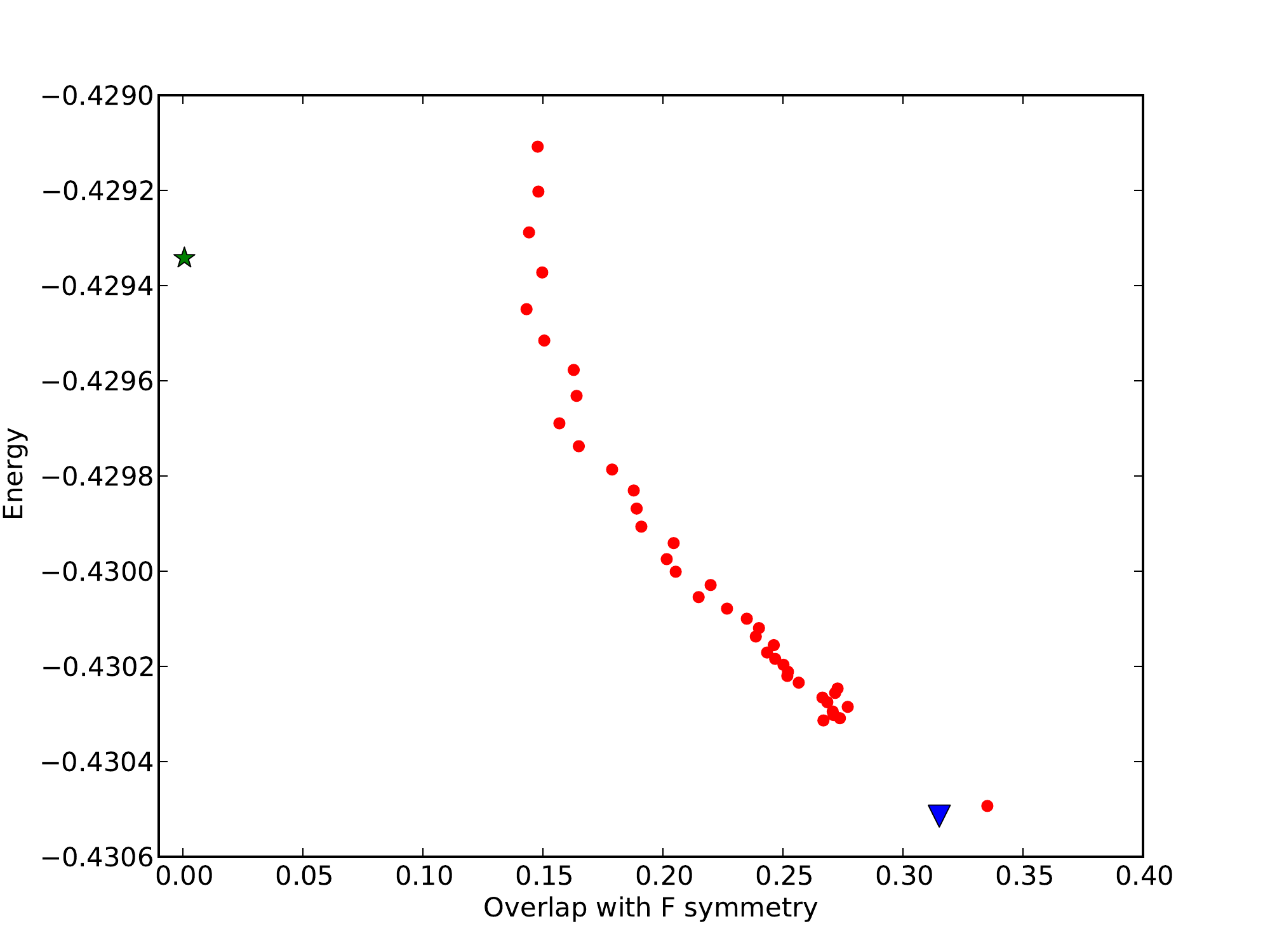}}
\caption{Spatial symmetry analysis of our state. 
(a) and (b) are the kagome space group representations associated with the pattern of Fig. \ref{fig:S.S}(b). 
%The $E_2$ pattern is allowed by symmetry given the existence of an $F_2$ pattern so that the symmetry breaking observed in \ref{fig:S.S}(b) arises purely from an emergence of an $F_2$ pattern.
(c) and (d) Correlation between the energy of different states and the amount of assymetry ((c): as defined by eqn.~\ref{eqn:GlobalAssymetry}; 
(d): as defined by eqn.~\ref{eqn:FAssymetry}). Red circles correspond to states
generated during part of an optimization, the green star to the Dirac spin liquid and the blue triangle to our most optimized state. For energies below the Dirac spin liquid,  improved energy is correlated with decreased total assymetry but increased 
strength of the $F_2$ pattern.}
\label{fig:symmetryanalysis}
%\label{fig:assymetryPlots}
%{\color{red} Another plot with a similar correlation for our measure of $U(1)\to Z_2$ symmetry breaking is necessary here. This could replace the site-assymetry plot.}}
\end{figure}

%\section{Symmetry analysis and relationship to a monopole creation operator}
Having established that our newly discovered state is very close to the Dirac state, we now turn to its symmetry breaking properties. To this end, we seek the space group representations that give the largest contribution to the pattern in Fig. \ref{fig:S.S}(b) which are periodic with at most a quadrupled unit cell. Such representations were constructed in Ref. \cite{Hermele2008} and labeled as $A_1$, $A_2$, $B_1$, $B_2$, $E_1$, $E_2$ for those related to the point group alone and $F_1$, $F_2$=$F_1\otimes A_2$, $F_3=F_1\otimes B_1$, $F_4=F_1\otimes B_2$ for those allowed by a doubling/quadrupling of the unit cell. The focus of Ref.~\cite{Hermele2008} was on the $F_1$ representation for the ``Hastings valence bond crystal'' states associated with the generation of mass of the Dirac fermions. However, the bond amplitudes plotted in Fig. \ref{fig:S.S}(b) are not of this representation. Instead, they are dominated by the $F_2$ and $E_2$ representations whose patterns are shown in Fig. \ref{fig:symmetryanalysis} (a) and (b). Remarkably, the symmetry of the $F_2$ pattern alone is the same as the symmetry of the $F_2$ and $E_2$ patterns. The $E_2$ pattern alone, however, has higher symmetry. Hence, the symmetry breaking observed here arises uniquely from a desire to form the $F_2$ pattern. 

The only time reversal symmetric alternative to the Hastings states, among instabilities of the Dirac fixed point identified in Ref. \cite{Hermele2008} is the spin singlet/nodal triplet ``$w$-monopole'' that is created by a complex operator $w_i$, $i=x,y,z$\cite{Hermele2008}. In \supacronym{} S-IV we show, following the transformation properties determined in Ref.~\cite{Hermele2008}, that the six dimensional vector $(\text{Re}\,w_x,\text{Re}\,w_y,\text{Re}\,w_z,\text{Im}\,w_x,\text{Im}\,w_y,\text{Im}\,w_z)^T$ transforms under the two three-dimensional representations $F_1$ and $F_2$. This remarkable coincidence allows us to conjecture that the $w$-monopole is responsible for the instability of the Dirac state observed in Fig.~\ref{fig:S.S}(b).  

%\begin{figure}[h]
%%\includegraphics[width=0.5\textwidth]{F2andE2.pdf}
%\includegraphics[width=0.2\textwidth]{kagome_e2.pdf}
%\hspace{0.2cm} \includegraphics[width=0.2\textwidth]{kagome_f2.pdf}
%\caption{Kagome space group representations associated with the pattern of Fig. \ref{fig:S.S}(b). The $E_2$ pattern is allowed by symmetry given the existence of an $F_2$ pattern so that the symmetry breaking observed in \ref{fig:S.S}(b) arises purely from an emergence of an $F_2$ pattern.}
%\label{fig:F2E2}
%\end{figure}

%{\color{red}The discussion below needs to be refocused on (1) how bond asymmetry correlates with energy and (2) and $U(1)\to Z_2$ symmetry breaking (i.e. pairing) correlates with energy. Does the measure of bond asymmetry requre the ADM? Can we do this with S.S?}
To further understand the role of the symmetry breaking, we measure the asymmetry of Fig.~\ref{fig:S.S} during part of an optimization run and correlate this with the observed energy.
We compare global asymmetry (removing the $E_2$ representation to remove the effects of twisted boundary conditions) defined as 
\begin{equation}
\label{eqn:GlobalAssymetry}
     {\mathcal O} = \sum_{\langle ij\rangle}\langle\Psi_0|\vec S_i \cdot \vec S_j|\Psi_0\rangle -  E_{2ij}^1 \langle\Psi_0|\vec S_i \cdot \vec S_j|\Psi_0\rangle 
\end{equation}
as well as 
the assymetry of the $F_2$ component  defined as 
\begin{equation}
\label{eqn:FAssymetry}
   {\mathcal O}_{F_2} = \sum_{\langle ij\rangle} F_{2ij}^1\langle\Psi_0|\vec S_i \cdot \vec S_j|\Psi_0\rangle
\end{equation}  
where $F^1_{2ij}=1/\sqrt{2N_b/3}$ on the solid thick bonds, $F^{1}_{2ij}=-1/\sqrt{2Nb/3}$ on the dashed thick bonds and zero otherwise with $N_b$ the number of nearest neighbor bonds. 
These are shown in Fig.~\ref{fig:symmetryanalysis} (c) and (d).
Interestingly, although we see a clear symmetry breaking, lower energy states have lower asymmetry, saturating at a value (triangle) that is small but still above the Wen Dirac state (star).  
%If the optimization was quenching out (but had not yet removed) the asymmetry or was optimizing toward a state that couldn't be capture by a projected mean field ansatz, we would expect to find lower energy states were also less asymmetric.
% We then, stochastically change the pairing function in small increments only accepting
%this stochastic change if our objective function indicates we have moved closer to a spin liquid state.  In this way, we move toward a 
%``nearby'' spin liquid. Of course, this procedure is not unique and 
%doesn't guarantee that the ``nearest'' spin liquid state will be found.
%Nonetheless, when we run this procedure, the spin liquid 
%state it approaches has the energy of the Dirac spin liquid 
%(and hence is presumably the Dirac spin liquid). This leads us to believe that
%the assymetry is fundamental to the gain in energy and 
%that a nearby spin liquid state at a similar energy that ``cleans up'' the 
%assymetry doesn't exist. 
%We find exactly the opposite feature, though, as the energy decreases for the overlap with the $F_2$ component instead finding lower energy states actually have higher overlap; from this we conclude that this type of spatial symmetry breaking is an important feature of our state. 
We find exactly the opposite behavior for the overlap with the $F_2$ component: lower energy states have greater overlap. This implies the symmetry breaking associated with the $F_2$ representation is an important feature of our state throughout much of the minimization process.

In addition to breaking the spatial symmetries of the lattice, the physical system also breaks the $U(1)$ symmetry of the Dirac state down to a $Z_2$ symmetry. To study this symmetry breaking we produced a series of runs without any pairing in the wave functions. 
We initialize twelve runs by taking the Dirac pairing function and multiplying each element by $(1+r)$ where $r$ is sampled in the interval $[0,X]$. Using 
$X=\{0.1,0.2,0.33\}$ we ran 12 simulations whose variation in initial energy was significant starting as high as $E=-0.375$ per site.  All simulations converged to $E=-0.4295$, the energy of the Dirac state.
%and significantly above the energy of the state presented in Fig.~\ref{fig:S.S}(b). 
From these results we conclude that the Dirac state lies at the bottom of a deep and wide region in orbital space and that the pairing of spinons is necessary to produce the state shown in Fig.~\ref{fig:S.S}(b).
Based on our results, it seems very likely that the Dirac fixed point is unstable to the formation of a stripe-like spin-liquid crystal phase. This conclusion rests on two assumptions: that fluctuations beyond those captured by the projected wave function do not restore the symmetry and that this symmetry breaking is not a finite size effect.  We have some indirect evidence supporting the latter assumption.  DMRG and exact diagonalization results indicate that a 4x4 unit cell cluster should be large enough to capture the qualitative physics of the system. In addition, we have looked at up to 8x8 unit cell clusters and can still find states with lower energy than the Dirac state.
%A priori we would find such a possibility surprising.  Either the two symmetry broken translational states would have to become degenerate leading to a ground state manifold that has a higher degeneracy than that predicted by the Dirac fixed point or different instabilities of the Dirac state would have to cross at finite $N$.  Although possible, the DMRG and exact diagonalization studies have suggested that a $4 \times 4$ unit cell cluster should be sufficiently large to capture the qualitative physics of the system.  In addition, we have looked at up to $8\times 8$ unit cell clusters, and find states with lower energy than the Dirac state. The instability is therefore unlikely a finite size effect.

%The other assumption is that fluctuations beyond those introduced by projection don't restore the symmetry and return the system to the Dirac fixed point. To address this possibility, we have explored a broader space of wave functions through fixed node Diffusion Monte Carlo and correlated product states. Though the resulting wave function has an energy of -0.433 J per spin significantly below -0.4305 J per spin of the projected BCS state it did not restore the symmetry. 

One way to directly address these assumptions would be to perform a PSG analysis on the relevant lower symmetry subgroup of the kagome lattice and use it to search for the state that projects to our state (we cannot do this directly because projection is a many-to-one mapping and cannot be inverted). Such an analysis would allow a determination of the finite size scaling of the symmetry breaking effects and provide a starting point for studying fluctuations about this phase in a low energy effective theory. This would also establish more directly the question of whether there is an energy gap (however, we expect such a gap because the wave vector of the symmetry breaking pattern connects the Dirac nodes of the Brillouin zone).

Given the lack of evidence in large-scale DMRG calculations for our state, it is unlikely to represent the true ground state. However, since it lies very
close by and involves a minimal loss of crystal symmetries, it seems likely to be a leading instability. In particular, small perturbations to the Hamiltonian could stabilize it suggesting it could be realized in nature. One promising class of materials are the Zn-Paratacamite family parameterized by Zn doping concentration $x$ with $x<1/3$. Unlike the structurally perfect kagome lattice of the $x=1$ Herbertsmithite member of the family, compounds with $x<1/3$, including clinoatacamite at $x=0$, break crystal symmetries and have distorted kagome layers\cite{SHLee2007} with precisely the distortion expected from the symmetry breaking of our state. Our results therefore motivate the study of single crystals of these materials and suggests that either the mysterious intermediate phase below $7 \text{K} <  T < 20\text{K}$ or the high temperature phase $T> 20\text{K}$ could still have spinons as low energy excitations that are delocalized along the ``rails'' of the distorted lattice. 

The most remarkable implication of our results is its suggestion that spin liquid crystal phases may be a common phenomena. Since any dimer state is the exact ground state at the mean field level\cite{Rokhsar1990}, projection must introduce quantum fluctuations that melt such crystalline phases, take the system through a succession of more symmetric phases until, in our case, it nearly reaches an isotropic phase. Such a picture has several implications for the DMRG calculations on the kagome lattice. It is known\cite{Yan2011} that small perturbations to the Hamiltonian in DMRG (boundary conditions, pinning fields, etc.) can enhance different states and lead to symmetry breaking. Exploring the class of perturbations that stabilize the symmetry breaking we observe here would help make a deeper connection between analytic Schwinger-fermion theory and DMRG. More interestingly, the observed DMRG state might be understood as a further instability of our state to a nematic spin liquid crystal, a state found recently on a triangular lattice model with ring exchange\cite{Grover2010}. A nematic spin liquid phase would be indistinguishable from a spin liquid phase on long cylinders that explicitly break rotational symmetry. Of course, it is also possible that such a putative nematic state melts into an isotropic $Z_2$ spin liquid. More generically, our results suggest that spin liquid crystal states are common in frustrated antiferromagnets and may be found using methods similar to ours as competitive ground states in many other systems.

%Perhaps the most remarkable implication of our results is the similarity between them and liquid crystals. The pattern we observe is clearly not a dimerization of the lattice as expected in a valence bond crystal. Through its one dimensional character, it resembles the smectic liquid crystal phase of a collection of rod-like polymers. Remarkably, an orientationally ordered spin liquid, analogous to a nematic liquid crystal, was also recently discovered by applying similar methods to ours to a triangular lattice Heisenberg model with ring exchange\cite{Grover2010}. As a speculation, it is also possible that the DMRG calculations are actually finding an orientationally ordered spin liquid, a possibility impossible to rule out at present since they are performed on long cylinders that explicitly break rotational symmetry. In such a case, a study of the melting of our stripes to a nematically ordered phase, a natural instability on symmetry grounds, would allow a direct comparison between the DMRG calculations and an analytic Schwinger-fermion theory based on our results. It is also possible that the true ground state in DMRG calculations has the symmetry of our state because the weak symmetry breaking we have discovered is very hard to notice. In any event, our results suggest that spin liquid crystal states are common in frustrated antiferromagnets and may be found using methods similar to ours as competitive ground states in many other systems.

\section{acknowledgements}
We acknowledge useful discussions with David Huse and Nandini Trivedi.  This work used the Extreme Science and Engineering Discovery Environment (XSEDE), which is supported by National Science Foundation grant number OCI-1053575 and the High Performance Computing Cluster at Case Western Reserve University.

%\bibliography{../../../Databases/BibTeX/kagomePSG}
%\bibliography{kagomePSG}
%

\end{document}